\documentclass[article,twoside]{revtex4}
\usepackage{color}
 \usepackage{amsbsy}
  \usepackage{bm}
 \usepackage{amsfonts}
\usepackage{amssymb}
\usepackage{amsmath}
\usepackage{graphics}
\usepackage{graphicx}
\begin{document}
\title{Effects of isospin imbalance on the chiral phase transition within a Chiral Dual Partner Model}

\author{R. M. Aguirre}
\affiliation{Departamento de Matematica, Universidad Nacional de
La Plata\\ and Instituto de Fisica La Plata,  CONICET\\ La Plata,
Argentina}

\begin{abstract}
The chiral symmetry is a property of the fundamental theory of the
strong interaction that is relevant for the hadronic physics. It
is expected that at sufficiently high temperature and matter
density, this symmetry becomes manifest. Within the Chiral Dual
Partner Model the chiral transformation is implemented in such a
way that a fermion mass term is allowed if the parity partner of
each baryon is included in the framework. This model is used here
to study possible manifestations of the chiral symmetry in dense
nuclear matter, assuming constant isospin fraction. It is found
that the onset of the odd parity baryons is associated with two
branches of thermodynamical instabilities, one of them leads to a
first order phase transition for temperatures below $T_c\simeq 11$
MeV. An analysis of these instabilities in the phase space is
given, and neutron star matter is considered as a special case.
\end{abstract}

\maketitle

\section{INTRODUCTION}

The earlier studies of the manifestations of the chiral symmetry
in a nuclear environment precede the formulation of a general
theory of the strong interaction
\cite{GellMann,Nambu,Nambu2,Levy,Dashen}. After the consolidation
of QCD as the fundamental theory, these studies were developed by
using effective models, due to the intricacies of this
formulation. The nontrivial structure of the vacuum, the running
coupling constant and the confinement mechanism turn almost
impossible a description of dense hadronic matter in terms of
quarks and gluons. These effective models try to extend the
abstract symmetries of QCD to the hadronic physics, restricting
their applicability to a range of phenomena characterized as the
low energy domain. Within such domain the relevant symmetry is
precisely the chiral symmetry.\\
One can find among the chiral effective models those that propose
a scheme of progressive approximations, but with a limited
applicability for the range of matter densities. The purpose of
other models is to capture the right dynamics by judiciously
choosing the degrees of freedom and their interactions. In such
case they usually rely on self-consistent approaches, as for
example the mean field approach (MFA) or similar, which allows to
extend the validity of the description to the densest systems
accessible to observation today. A reduced number of examples are
given by \cite{Matsui,Glendenning,Detar,HatsudaP,Jido,Papazoglou,Mishustin}.\\
The standard prescription for the chiral transformation require
massless fermion fields, therefore the particles with a finite
mass must obtain it dynamically. The coupling with a scalar meson
field having a nonzero vacuum expectation value is the usual
mechanism for this purpose. However, it has been pointed out
\cite{Detar,HatsudaP,Jido} that the implementation of the chiral
symmetry through an alternative prescription, known as
\textit{mirror}, allows the existence of a mass term for the
fermion fields. This possibility was used to introduce a mass
parameter $m_0$ which could be interpreted as a contribution from
the quark structure that can not be accounted for solely by the
hadronic interaction. It is expected that at high
density/temperature the dynamical mass decrease to its minimum
value $m_0$, this is the signal of the chiral restoration in a
hadronic medium. The mirror assignment for the chiral symmetry is
accomplished by introducing a complementary parity state for each
baryonic degree of freedom, for this reason this framework is
usually known as the Chiral Dual Partner Model (CDPM). \\
The CDPM has been extended from its early formulation to include a
variety of issues, as for instance the phase diagram in an SU(2)
treatment \cite{Zschiesche}, axial vector mesons \cite{Gallas},
neutron stars \cite{DexSchrammZschi}, hyperons
\cite{DexSteinh,Steinh,Mukherjee,Sasaki,Mukherjee2,Motornenko,FragaSchaff,Steinh3},
the delta isobar \cite{Takeda,MarczeRedlich2,MarczeRedlich},
fluctuations of the conserved charges
\cite{Marczenko1,Marczenko2,Koch,Marczenko3}, isospin imbalanced
nuclear matter \cite{Motohiro,EserBlaiz2}, corrections by the
method of functional renormalization group \cite{Weyrich,Tripolt},
and Dirac sea effects \cite{EserB}. Other approaches proposed an
hybrid composition of the CDPM and quark interactions
\cite{Minamikawa,GaoMinamik,Kong,GaoKong}.

 The models aimed to describe the dense
hadronic medium are subject to a variety of phenomenological
constraints. In addition to the vacuum low lying mass spectrum,
they should also give an acceptable explanation for the gross
features of the atomic nuclei and to adjust the properties of the
homogeneous nuclear matter at the saturation baryonic density
$n_0=0.15$ fm$^{-3}$, such as binding energy, compressibility,
symmetry energy and its slope parameter. The effective nucleon
mass at the saturation density $M^*$ plays a special role, it has
not been measured precisely but it is expected to vary within a
restricted domain. The scalar interaction produces a decrease of
this effective mass with the density, which in turn, affects the
spin-orbit coupling for the atomic nuclei. It has been found that
the larger $M^*$ the weaker spin-orbit term \cite{Koepf},
therefore the reproduction of the energy levels for the closed
shell nuclei of medium mass imposes a restriction on $M^*$.
 Furthermore, the increasing variety and precision of the
astronomical observational data provides useful information, such
as the mass-radius relation for neutron stars and the tidal
deformability of a binary system, which involves the high density
equation of state and must be properly taken into account.

The isospin imbalance is a common feature of nuclear dense systems
which manifests, for instance, in the predominance of the neutron
over the proton fractions. In models of meson exchange the nuclear
isospin density is the source for the isovector mesons $\rho
(780)$ and $a_0 (980)$, which modify the energy spectrum and the
effective mass of the fermions, and consequently their chemical
potentials. In the CDPM the chiral phase transition is driven by
the onset of the parity partners of the proton and neutron, which
is determined by their chemical potentials. Therefore, in a system
with conserved isospin configuration it is expected that the
coupling to the isovector mesons must have sizeable effects on the
chiral transition. There is only a reduced number of publications
that considered the inclusion of the $a_0 (980)$ meson within the
CDPM \cite{Kong,GaoKong}, but they impose a transition to quark
matter at a relatively low baryonic density $n_B/n_0=2$. Therefore
it is possible that such effects were not detected.\\
Another interesting reason to study strong interacting systems
with isospin imbalance is that this would be the bridge between
the hadronic phenomenology and the results obtained by QCD lattice
simulations  \cite{Son}. These simulations have technical
difficulties to treat finite baryonic density, although they are
viable in the presence of a finite isospin chemical potential
$\mu_I$ \cite{Alford}. Under such conditions the ground state
obtained for low $\mu_I$ contains a pion condensed phase,  and
this condensate modifies the dynamical properties of the fermions
\cite{Bedaque}.

This work is devoted to the study of possible manifestations of
the chiral symmetry in a dense hadronic environment with isospin
imbalance. The SU(2) CDPM with the addition of a scalar isovector
meson is used for this purpose. The remaining of this work is
organized as follows, in the next section a brief description of
the model and the procedure used for fix its parameters is given.
The main results are presented and discussed in Sec.
\ref{Results}, they correspond to homogeneous matter with a
constant isospin fraction and to neutron star matter as a special
case. The conclusions are drawn in Sec. \ref{Colussions}.

\section{THE MODEL}

In this work an extended version of the CDPM suited to study
isospin asymmetric matter is used. Therefore, as in previous
investigations
\cite{HatsudaP,Minamikawa,Motohiro,Motornenko,DexSchrammZschi,DexSteinh,EserBlaiz2,FragaSchaff,GaoMinamik,GaoYan,GaoYuan,GaoKong},
I include the vector isovector $\rho (770)$ meson. Furthermore, an
explicit coupling between the nucleons and the scalar isovector
meson $a_0 (980)$ is considered. Its effects have been usually
disregarded, although a few investigations \cite{Kong,GaoKong}
have studied its role in the star matter at zero temperature.
Thus, with exception of this addition, the usual procedure for the
CDPM is applied here. For the sake of completeness it is
summarized in the following. The Lagrangian density for the
fermions is written in terms of the isospinor fields $\Psi_1$ and
its chiral partner $\Psi_2$, which have opposite parity
\begin{equation}{\cal L}_N=\bar{\Psi}_1 i\not\!\!D \Psi_1-g_1 \bar{\Psi}_1
\Sigma \Psi_1+ \bar{\Psi}_2 i\not\!\!D \Psi_2-g_2 \bar{\Psi}_2
\Sigma^\dag \Psi_2-m_0 \left( \bar{\Psi}_1 \gamma_5 \Psi_1-
\bar{\Psi}_2 \gamma_5 \Psi_2\right)\end{equation}
where $\Sigma=\sigma+ \bm{\tau} \cdot \bm{\zeta}+i \gamma_5
\bm{\tau} \cdot \bm{\pi}$ collects the scalar and pseudosclar
vertices, and $D_\mu=\partial_\mu+i\,g_w \omega_\mu+i\, g_r
\bm{\tau} \cdot \bm{\rho}_\mu$ includes the vector meson
couplings. Despite the mass term, this construction is chiral
invariant under the mirror prescription \cite{Detar,Jido},
\[\left[Q^a_V,\Psi_k\right]=-t^a\,\Psi_k,\; \left[Q^a_A,\Psi_k\right]=-t^a\,\eta_k\,\Psi_k \]
for $Q_V,\,Q_A$ the vector and axial charges, $t_a=\tau_a/2$, and
$\eta_k=(-1)^{k+1}$.\\
The meson sector is decomposed as the sum of the kinetic and
interaction terms
\begin{equation}{\cal L}_{\text{kin}}=\frac{1}{2}\partial_\mu \sigma\, \partial^\mu \sigma+\frac{1}{2}\partial_\mu \bm{\zeta}\cdot \partial^\mu \bm{\zeta}
-\frac{1}{4}W_{\mu \nu} W^{\mu \nu}-\frac{1}{4}R_{\mu \nu}
R^{\mu\nu}+\frac{1}{2}m_w^2 \,\omega_\mu
\omega^\mu+\frac{1}{2}m_r^2 \,\bm{\rho}_\mu\cdot
\bm{\rho}^\mu\end{equation}
\begin{eqnarray}{\cal L}_{\text{int}}&=&\frac{C_0}{4}\,\text{Tr}\left(\Sigma^\dag
\Sigma\right)-\frac{C_1}{16}\,\left[\text{Tr}\left(\Sigma^\dag
\Sigma\right)\right]^2-\frac{C_2}{8}\,\text{Tr}\left(\Sigma^\dag
\Sigma\right)^2+h
\left[\text{Tr}\left(\Sigma+\Sigma^\dag\right)/2-f_\pi\right]\nonumber
\\
&&-\sum_3^4 \frac{\alpha_n}{2^n n!}\left[\text{Tr}
\left(\Sigma^\dag\Sigma\right)/2-f_\pi^2\right]^n-\frac{\alpha_0}{8}\left(\omega^4+\rho^4+6
\omega^2 \rho^2\right)+C_3 \label{LagVec}\end{eqnarray}
The scalar mesons interaction, in the first line, resembles that
of the Linear sigma model (LSM) \cite{Levy,Lenaghan}, but it is
complemented here by polynomial terms of order six and eight,
shown in the second line. Such completion has been a common
practice for the use of nuclear chiral models
\cite{Floerch,Drews1,Drews2,FragaHipp,FragaMata}, which has been
extended to the CDPM \cite{FragaSchaff}. In the vector meson
sector, a self-interaction is included in Eq. \ref{LagVec}, using
the convention $\omega^{2n}=\left(\omega_\mu \omega^\mu\right)^n$
and $\rho^{2n}=\left(\bm{\rho}_\mu \cdot \bm{\rho}^\mu\right)^n$.
Nuclear models have introduced this class of vertices to fulfill
phenomenological requirements, particularly in relation to the
isospin response. It has also been previously associated with the
CDPM, as for instance in \cite{FragaSchaff}.\\
In the LSM the coefficients of the scalar potential are usually
chosen to adjust the masses of the lightest mesons. This can be
done by expanding the scalar meson fields around their vacuum
expectation values, $\sigma=s_0+S$,
$\zeta_a=z_0+Z_a$, $\pi_a=\pi_0+P_a$, with
$s_0=f_\pi$, $z_0=\pi_0=0$ and defining
\[m_F^2=\frac{\delta^2 {\cal L}}{\delta F\, \delta F^\dagger}(S=0,\bm{Z}=0,\bm{P}=0),\;\;F=S,\,
Z_a,\, P_a\]
Thus, the meson masses are known even if the
calculations were performed at the tree level.\\
The same decomposition can be applied at finite density and
temperature $\sigma=s+\delta\sigma$, $\zeta_a=z \delta_{3
a}+\delta\zeta_a$, $\pi_a=\delta\pi_a$, $\omega_\mu=w\, g_{0
\mu}+\delta\omega_\mu$, $\rho^a_\mu=r \,\delta_{a 3}g_{0
\mu}+\delta\rho^a_\mu$, replacing vacuum by in medium expectation
values.  Assuming homogeneous and isotropic matter these
expectations values behave as uniform background fields. In this
work the mean field approach (MFA) is used, which is obtained by
neglecting the meson fluctuations. Within this approach the baryon
mass matrix in the isotopic space of the chiral partners is given
by
\begin{eqnarray} {\cal M}=\left( \begin{array}{cc}g_1(s+I_a z)& m_0
\gamma_5\\-m_0 \gamma_5&g_2(s+I_a z)
\end{array}\right),
\end{eqnarray}
where $I_a=\tau^3_{aa}$. Furthermore the meson fields matrix reduces to
$\Sigma=\text{diag}\left(s+z,s-z\right)$.\\
To obtain the mass eigenstates, which will be identified as the
nucleons and their chiral partners, a linear transformation is
proposed
\begin{eqnarray}
\left( \begin{array}{c}
\Psi_1^a\\\Psi_2^a\end{array}\right)=\left(
\begin{array}{cc}\cos\theta&-\gamma_5 \sin\theta\\\gamma_5 \sin\theta &\cos\theta
\end{array}\right)\, \left(\begin{array}{c}
N_1^a\\ N_2^a\end{array}\right)
\end{eqnarray}
By choosing $\tan\theta=m_0/G_1 (s+I_a z)$, the eigenvalues of the
mass matrix
\begin{equation}M_i=G_2 (s+I_i z)\,J_i +\sqrt{m_0^2+G_1^2 (s+I_a z)^2},\label{NuclMass}\end{equation}
 are obtained, where $G_1=(g_1+g_2)/2$,
$G_2=(g_1-g_2)/2$, $I_i=(-1)^{i+1}$, and $J_i=1$ for $i=1, 2$,
$J_i=-1$ for $i=3, 4$. The indexes are assigned by the convention
$i=1,\,2$ for the proton ($p$) and neutron ($n$) respectively,
while the odd parity partners $p^*$ and $n^*$ are identified by
$i=3,\, 4$ respectively.

Within the approach just described, the grand potential per unit
volume of the system can be written as
\[\Omega(s,z,w,r,T,\mu_k)=\sum_{i=1,4} \Omega_i(T,M_i,\mu_i)-U(s,z)-V(w,r), \]
the sum extends over isotopic and parity components, and
\begin{equation}\Omega_i=-\frac{2}{\beta}\int\frac{d^3p}{(2
\pi)^3}\left[\ln\left(1+e^{-\beta \alpha_i} \right)+
\ln\left(1+e^{-\beta \bar{\alpha}_i} \right)\right],\label{GrandP}
\end{equation}
where $\beta=1/T$, $\alpha_i=\sqrt{p^2+M_i^2}+ g_w w+ g_r I_i
r-\mu_i$, and $\bar{\alpha}_i=\sqrt{p^2+M_i^2}- g_w w- g_r I_i
r+\mu_i$. Since the conservation of both the baryon number and the
isospin composition are analyzed in this work, one must consider
two chemical potentials $\mu_B$ and $\mu_I$, respectively. The
particle chemical potentials are related to these, i.e.
$\mu_1=\mu_3=\mu_B-\mu_I$, $\mu_2=\mu_4=\mu_B+\mu_I$. Furthermore,
the meson potentials in the MFA are given by
\[V=\frac{1}{2}\left[(m_w w)^2+(m_r r)^2 \right]+\frac{\alpha_0}{4}\left(w^4+r^4+3 w^2 r^2 \right) \]
\[U=\frac{C_0}{2}\left(s^2+z^2 \right)-\frac{C_1}{4}\left(s^2+z^2 \right)^2-\frac{C_2}{4}
\left(s^4+6 s^2 z^2+z^4 \right)+h (s-f_\pi)-\sum_{n=3,4}
\frac{\alpha_n}{2^n n!}\left(s^2+z^2-f_\pi^2\right)^n+C_3.
\]
The meson expectation values are determined by the extremum
condition for $\Omega$, leading to the equations
\[s\,\left[-C_0+C_1 \left(s^2+z^2\right)+C_2 \left(s^2+3 z^2\right)+
\sum_{n=2,3}\frac{\alpha_{n+1}}{2^n
n!}\left(s^2+z^2-f_\pi\right)^n\right]-h+\sum_{i=1,4}N_i
\frac{\partial M_i}{\partial s}=0,\]
\[z\,\left[-C_0+C_1 \left(s^2+z^2\right)+C_2 \left(z^2+3 s^2\right)+
\sum_{n=2,3}\frac{\alpha_{n+1}}{2^n
n!}\left(s^2+z^2-f_\pi\right)^n\right]+\sum_{i=1,4}N_i
\frac{\partial M_i}{\partial z}=0,\]
\[w \left[m_w^2+\alpha_0 \left(w^2+3 r^2\right)\right]-g_w n_B=0,  \]
\[r \left(m_r^2+3 \alpha_0 w^2\right)-g_r n_I=0.  \]
The total barionic density is the sum of the partial contributions
$n_B=\sum_i n_i$, as well as the isospin asymmetry $n_I=\sum_i I_i
n_i$, where
\[ n_i=2\,\int \frac{d^3p}{(2 \pi)^3}\left[n_F(T,\mu_i)-\bar{n}_F(T,\mu_i) \right],\]
\[N_i=2\,\int \frac{d^3p}{(2 \pi)^3}\,\frac{M_i}{E_i}\left[n_F(T,\mu_i)+\bar{n}_F(T,\mu_i) \right],\]
and $n_F, \,\bar{n}_F$ stand for the canonical Fermi distribution
functions for particles and antiparticles, and
$E_i=\sqrt{p^2+M_i^2}$. Finally, the energy per unit volume is
obtained by the thermodynamical relation ${\cal E}=\Omega+
T\,{\cal S}+\sum_i \mu_i n_i$, where the entropy density is
obtained as usual by ${\cal S}=-\partial \Omega/\partial T$.

This model has several free parameters, some of them can easily be
related to the meson phenomenology. As in the LSM the constants
$C_0, C_1, C_2$ are related to the meson masses in the MFA as
\[C_0=\left(m_\sigma^2-3 m_\pi^2\right)/2, \;C_1=\left(m_\sigma^2-
m_\zeta^2\right)/2\,f_\pi^2, \; C_2=\left(m_\zeta^2-
m_\pi^2\right)/2\,f_\pi^2.\]
Furthermore $h=f_\pi m_\pi^2$ is chosen in order that the vacuum
expectation value of the scalar fields be $s_0=f_\pi=93$ MeV,
$z_0=0$. Whereas $C_3$ is determined by requiring that the
pressure $P=-\Omega$  vanishes at zero temperature and density. On
the other hand, taking the nucleon mass $M_1^0=939$ MeV and
choosing the chiral partner as $M_2^0=1530$ MeV, one obtains
$G_2=\left(M_1^0-M_2^0\right)/2 \,f_\pi$.

 The remaining constants are used to adjust the
nuclear matter phenomenology at $T=0$ and normal density
$n_0=0.15$ fm$^{-3}$, and to obtain compatibility with
observational data on compact stars. The following data on
symmetric nuclear matter is used,
\begin{equation}P(n_0)=0,\;\frac{{\cal E}(n_0)}{n_0}=923\,\text{MeV,}\;
E_{\text{sym}}=\frac{n_0}{2}\,\frac{\partial^2{\cal E}}{\partial
n_I^2}=30 \,\text{MeV,}\label{NucMatt1}\end{equation}
\begin{equation}L_{\text{sym}}=3\,n_0\,\frac{d
E_{\text{sym}}}{d n}=70\,\text{MeV,}\;K=9\,n_0\,\frac{d\mu_B}{d
n}=260\,\text {MeV}.\label{NucMatt2}\end{equation}
corresponding to the condition of minimum energy per particle,
binding energy, symmetry energy and its slope, and the
compressibility, respectively. Some of the chosen values are of
qualitative nature since they have not been precisely measured
yet. Furthermore, there is some tension between the commonly
accepted values and the recent analysis of the experimental data
on the symmetry properties $E_{\text{sym}}$ and $L$
\cite{Horowitz,Reed,Reed2}.
\\
Similarly, the nucleon effective mass in a dense medium $M^*$ is
poorly known. However, a possible range of variation has been
established, mainly due to nuclear structure estimations. In this
work  the range $0.65 < M_1^*/M_1^0 < 0.8$ is explored for the
nucleon mass, Eq. \ref{NuclMass}, at the normal density. For the
same reason, the sigma meson mass is varied within the range $200$
MeV $< m_\sigma< 500$ MeV.

\begin{figure}
    \centering
    \includegraphics[height=0.4\textheight] {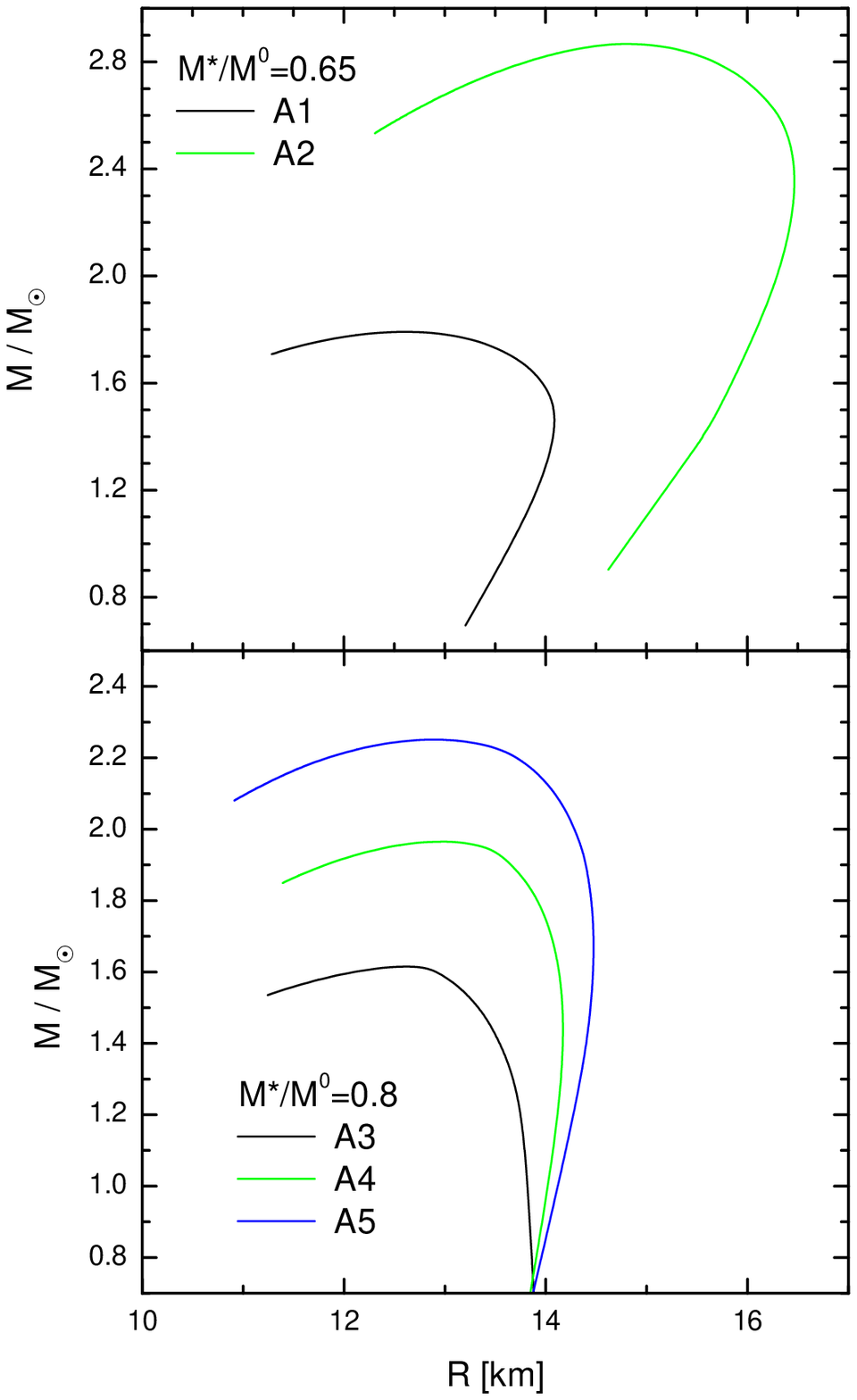}
    \caption{The mass-radius relation for neutron stars for different parametrization.
    The upper (lower) panel corresponds to the cases with an
    effective nucleon mass fixed at $M_1^*/M_1^0=0.65$ ($M_1^*/M_1^0=0.8$).}
    \label{Star}%
\end{figure}

\section{RESULTS AND DISCUSSION}\label{Results}

In first place I try to tune up the undetermined parameters in
order to reproduce the properties of nuclear matter just
mentioned. For this purpose the selected domain in the
$M_1^*-m_\sigma$ plane is explored. It is found that for a given
value of $M_1^*$, these conditions can not be adjusted for the
lowest value of $m_\sigma$. On the other hand, the parameter
$\alpha_0$ of the $\omega-\rho$ mixing decreases with $m_\sigma$,
and it becomes negative for $m_\sigma\sim 400$ MeV. This in turn,
breaks down the equation of the $\omega$ expectation value for
sufficiently high densities. Aiming to the study of neutron stars,
densities up to $n/n_0=10 $ are considered here. Therefore a more
restrictive condition $300 \, \text{MeV,}\leq m_\sigma \leq 400$
MeV must be imposed. It must be mentioned that small variations of
the numerical values taken in Eqs. \ref{NucMatt1}-\ref{NucMatt2},
produce continuous small
changes in the mentioned results.\\
Applying the criterion just stated, a reduced set of parameters is
selected to study star matter, that is nuclear matter stable
against beta decay, and electrically neutral. As explained in more
detail in the next section, a cloud of free electrons must be
considered to fulfill these conditions. The equations of state
obtained are introduced in the Tolman-Oppenheimer-Volkoff equation
for the structure of a non-rotating compact star. The results for
the mass-radius relation of the star are shown in Fig. \ref{Star}.
In the upper panel the results for $M_1^*/M_1^0=0.65$ and two
cases with different $\sigma$  meson masses $m_\sigma=300$ MeV
(A1), and $m_\sigma=391.4$ MeV (A2), corresponding to the vector
meson mixing $\alpha_0=1241.4$ and $0$, respectively. In the lower
panel, the case $ M_1^*/M_1^0=0.8$ is presented for three
different values $m_\sigma=300$ MeV (A3), $m_\sigma=400$ MeV (A4),
and $m_\sigma=431$ MeV (A5). They correspond to $\alpha_0=437.8,
\, 60.1$, and $0$ respectively. As a first conclusions, it is
found that the greater the value of $ M_1^*$, the wider the range
of admissible $m_\sigma$. In addition, the maximum mass of the
star corresponds to the higher $m_\sigma$. Since there is
empirical evidence of compact stars with a gravitational mass
greater than two solar masses, we keep only the parametrizations
A2 and A5, which are specified in Table \ref{Table1} with the new
labels A and B respectively.

\begin{table}[b]
\begin{tabular}{l|c|c|c|c|c|c|c}
Set & $m_0$ [MeV] & $G_1$ & $g_w$ & $g_r$ & $\alpha_3$
[MeV$^{-2}$] & $\alpha_4$ [MeV$^{-4}$] & $m_\sigma$ [MeV] \\
\hline A & 493.72 & 12.17 & 11.76 & 9.61 & -0.1146$^2$ &
0.0318$^4$ & 391.4 \\
\hline B & 622.69 & 11.46 & 8.27 & 7.39 & -0.1034$^2$ & 0.049$^4$
& 431.1
\end{tabular}
\label{Table1} \caption{The two sets of parameters used in this
work. For both cases is $G_2$=-3.18.}
\end{table}

\subsection{Neutron Stars}
 Different
versions of the CDPM were used previously to study neutron stars,
as for instance in
\cite{HatsudaP,DexSchrammZschi,PagliaraTolos,DexSteinh,FragaSchaff,MarczeRedlich,MarczeRedlich2,Steinh,Yuan,
Minamikawa,GaoMinamik}. Some of these works consider the presence
of hyperons \cite{Steinh,Yuan} or the transition to deconfined
quark matter \cite{Minamikawa,GaoMinamik} or both of them, hence
they are not directly comparable to the present results. In other
cases \cite{DexSchrammZschi,PagliaraTolos,DexSteinh}, the mean
field scheme was unable to reconcile the constraint of a
gravitational mass greater than two solar masses. However, the
addition of different physical considerations allows to fulfill
this condition. For instance, the inclusion of vacuum effects in
\cite{PagliaraTolos}, or a sixth order polynomial self-interaction
of the scalar mesons in \cite{MarczeRedlich,MarczeRedlich2}. It is
remarkable that such polynomial self-interaction, with an order
higher than four, seems to be a key ingredient to reconcile the
low density nuclear phenomenology and the astronomical data.\\
In this simplified treatment certain complemental degrees of
freedom, such as hyperons and free quarks, are not taken into
account. Several publications using the CDPM are devoted to this
specific issue
\cite{Steinh,DexSteinh,Mukherjee,Sasaki,Mukherjee2,Motornenko,FragaSchaff,FragaMata,Steinh3,GaoYuan}.
The presence of hyperons could have significant effects on the
structure of neutron stars, giving rise to the effect known as
hyperon puzzle, but this is not the case for the CDPM. In fact,
some calculations \cite{FragaSchaff} have shown a negligible
effect on the mass-radius relation of neutron stars. \\
\begin{figure}[b]
    \centering
    \includegraphics[height=0.4\textheight] {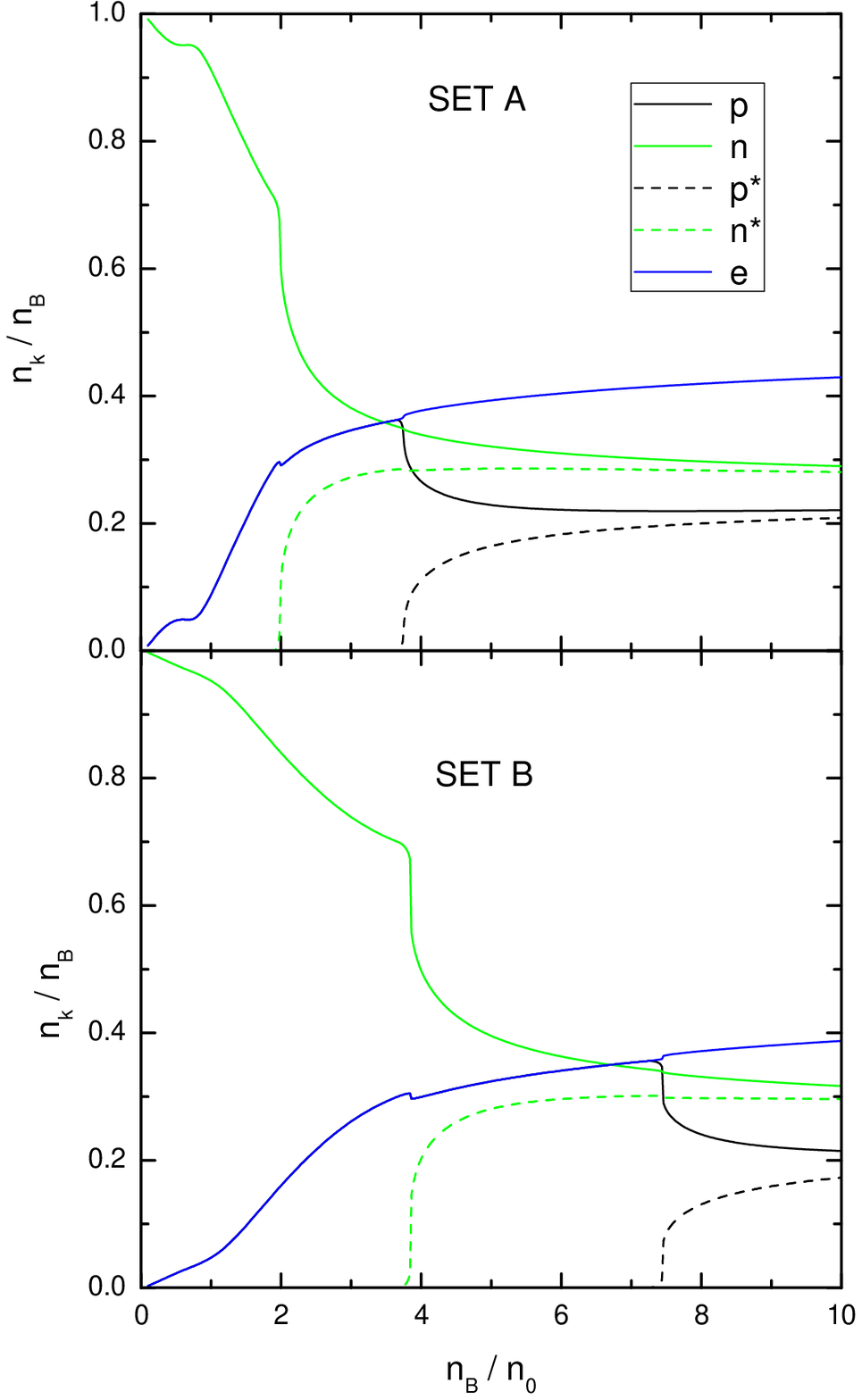}
    \caption{The population of different particles in the stellar
    environment, in terms of the total density.
    The upper (lower) panel corresponds to the parametrization case A (case B).}
    \label{StarPop}%
\end{figure}

\begin{figure}[h]
    \centering
    \includegraphics[height=0.4\textheight] {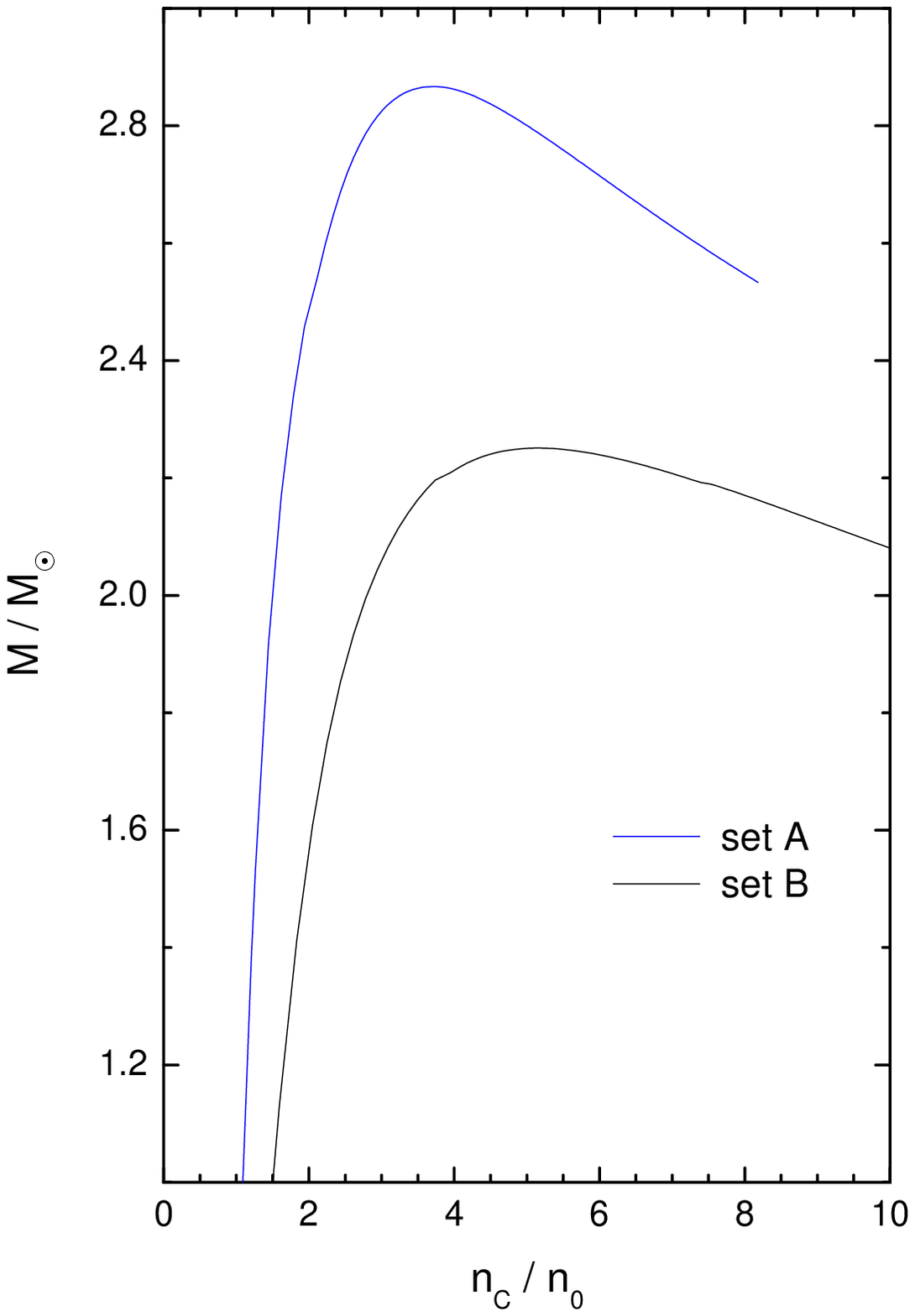}
    \caption{The gravitational mass of neutron stars in terms of its central density.}
    \label{StarCentral}%
\end{figure}

For the sets A and B is $\alpha_0=0$, so the mean values of the
vector mesons are simply given by $w=g_w n_B/m_w^2$ and $r=g_r
n_I/m_r^2$. At zero temperature, the results for the pressure and
the energy density are given by
\[{\cal E}=\frac{1}{8 \pi^2}\sum_{i=1-4, e} \left[p_{F i} E_{F i}\left(M_i^2+2\,p_{F i}^2 \right)
-M_i^4 \ln\left(\frac{p_{F i}+E_{F i}}{M_i}
\right)\right]+V(w,r)-U(s,z), \]
\[P=\sum_i \mu_i n_i+\,\mu_e n_e-{\cal E},  \]
in terms of the Fermi momentum $p_{F i}$ and energy $E_{F i
}=\sqrt{p_{F i}^2+M_i^2}$.  The conservation of the baryon number
density and electric charge are reflected by the equations
$n_B=\sum_{i=1-4} n_i$ and $0=n_1+n_3-n_e$ respectively. As usual
the number density for a particle is given by $n_i=p_{F i
}^3/3\,\pi^2$. The associated chemical potentials are related to
the particle chemical potentials by $\mu_B=\mu_2=E_{F 2}+g_w w-g_r
r$, $\mu_Q=-\mu_e=E_{F e}$. The stability against beta decay
imposes $\mu_2-\mu_e=\mu_1$, and $\mu_2=\mu_4\;\mu_1=\mu_3$ if the
dual partners are present. The onset of the dual partners
takes place at densities for which $\mu_2=M_4,\; \mu_1=M_3$.\\
%****
The assumption of homogeneous matter is broken at very low
densities, giving place to the formation of finite nuclei. For
this reason, within such domain, the results obtained are joined
in a continuous way and replaced by the equation of state given in
\cite{Baym}. This is the composite input used for the
Tolman-Oppenheimer-Volkoff equation
\begin{eqnarray}
\frac{dP}{dr}&=&- \,\frac{1}{r}\, [{\cal E}(r)+P(r)]\,\frac{{\cal
M}(r)+4 \pi
r^3 P(r)}{r-2 {\cal M}(r)} \, , \nonumber \\
{\cal M}(r)&=&\int_0^r 4 \pi \, {r'}^2 \, {\cal E}(r') \, dr' \, .
\nonumber
\end{eqnarray}
to obtain the results shown in the next figures. Units for which
$c=1, \,G=1,\, \hbar=1$ have been used.\\
In Fig.\ref{StarPop} the population of the different particles are
shown for a wide range of densities.  As the density is increased
the parity partner $n^*$ emerges earlier than the $p^*$. This is a
consequence of two facts,  beta stability and a non-zero mean
field value $z>0$. The first condition requires that $\mu_1 <
\mu_2$, and the second one leads to $M_3 > M_4$. Since $\mu_2$
increases with the density and from the inequalities $M_3
> M_4 \geq \mu_2 > \mu_1$, it follows that the threshold $\mu_2=M_4$ is reached before
than $\mu_1=M_3$. For the set A this happens at a notably low
density, slightly lesser than $n_B/n_0 \simeq 2$, while for the
set B this fact is delayed until $n_B/n_0 \simeq 4$. This feature
can be explained by the lower value of $m_0$ for the set A, see
Table \ref{Table1}, which makes the baryonic dynamics more
reactive to the in-medium effects.\\
The threshold of the $n^*$ is marked by a strong decrease of the
neutron population. The proton and electron curves coincide for
densities below the threshold of $p^*$, as required by the charge
neutrality condition. At relatively large densities, the parity
partners are almost equally populated. The combined analysis of
Figs. \ref{StarPop} and \ref{StarCentral}, the latter exhibiting
the gravitational mass of a star in terms of its central density,
shows that the stars having the standard mass $M/M_\odot = 1.4$
are composed solely by nuclear matter. The odd parity $n^*$ enters
in the composition of the stars having masses around the maximum
value, and for the set A a scarce fraction of $p^*$ can also be
found.

\begin{figure}
    \centering
    \includegraphics[height=0.4\textheight] {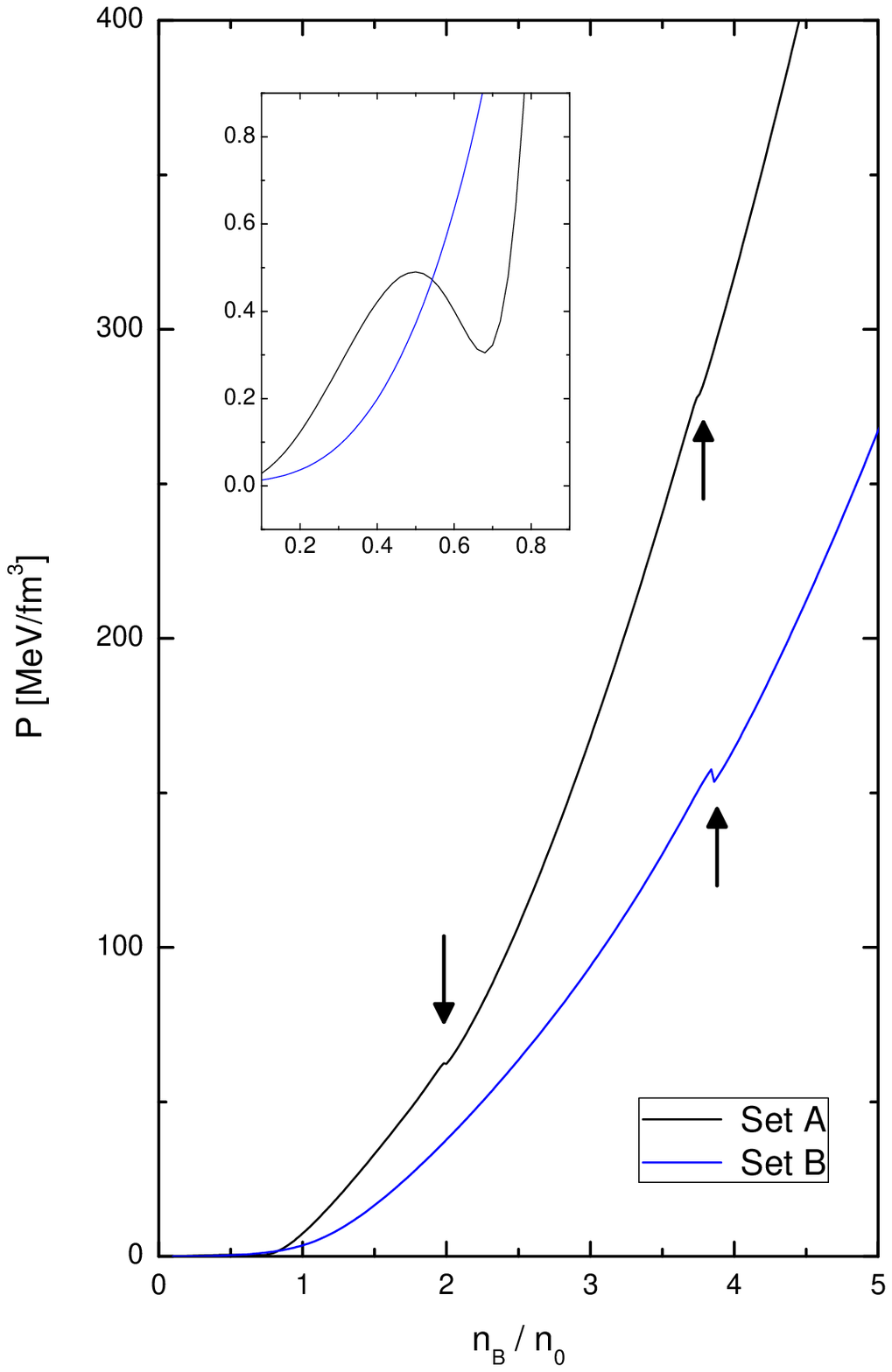}
    \caption{The pressure inside a neutron star in terms of the total density.
    The arrows indicate the position of the $n^*$ threshold, and of the $p^*$ only for the set A.}
    \label{StarEoS}%
\end{figure}

The onset of the dual partners is associated to thermodynamical
instabilities, as shown in Fig. \ref{StarEoS}. Near their
activation points the pressure becomes locally non-monotonous,
with the exception of  $n/n_0\simeq 3.5$ for the set A where the
only manifestation is a slight change in the slope of the
pressure. In such case an abrupt change in the compressibility of
the system is expected. The small insert in this figure, shows in
detail the low density region. For the set B a monotonous
increasing pressure is obtained, while for the set A a van der
Waals-like equation of state undoubtedly indicates the liquid-gas
phase transition. This transition is a common feature of the
nuclear matter, however it is unusual in the neutron star matter.

At this point it is useful a contrast with the more recent
analysis for the massive pulsar PSR J0740+6620 having a measured
mass $M/M_\odot=2.08\pm 0.07$. Its radius has been estimated
within the range $11.7 <R\, [\text{ km}]<14.25$ \cite{Miller}. The
result A, with a prediction of $R\simeq 16.3$ km is clearly  out
of the tolerance range, while the value $R\simeq 14.1$ km given by
the set B, is acceptable although greater than the preferred
estimation $R\simeq 12.8$ km. Another interesting piece of
evidence corresponds to the compact object PSR J0030+0451, which
was first identified as a neutron star with mass $M/M_\odot=1.4$
and a radius $R\simeq 13$ km. A recent revision \cite{Luo2024} has
estimated its mass and radius as $1.3< M/M_\odot<1.6$, and $11.7
<R\, [\text{ km}]<12.9$. Within such mass domain the predictions
of the present model are $R\simeq 15.7$ km (A), and $R\simeq 14.4$
km (B). However, another interpretation of the same observational
data \cite{Vincig} considers that $M/M_\odot=1.7$, $R\simeq 14.5$
km, which is compatible with the result given by the set B.\\
The parametrization set A admits heavy stars  $2 < M/M_\odot <
2.8$, but having relatively large radii $R > 16$ km, which is in
tension with presently accepted values. For this reason the
results corresponding to the set A will be postponed, and only the
set B is considered for the remaining of this work.

\subsection{THE CHIRAL TRANSITION IN ISOSPIN IMBALANCED NUCLEAR MATTER}
Up to this point an effective lagrangian has been defined that
takes account of the vacuum masses of nucleons and their parity
partners, as well as the masses of the pion and a0(980) meson. For
the $\sigma$ meson mass a value consistent with the known
phenomenology has been chosen. In addition, it is able to adjust
the following properties of dense matter: saturation density,
binding energy, compressibility, symmetry energy and its slope
parameter. The effective nucleon mass at the normal nuclear
density is slightly high, but it falls within the range of
acceptable values. Finally, the prediction for the mass-radius of
a neutron star passes the test of the results usually taken as
standards. Thus the fitting of masses of the low lying mesons has
been incorporated to the general scheme of fixing parameters, this
criterium is similar to that used in
\cite{Zschiesche,MarczSasaki}. This is a common practice within
the LSM which is not adequately considered in the applications of
the CDPM.

In this section the isospin composition of matter is taken into
account by introducing the isospin fraction
$\text{w}=(n_2+n_4-n_1-n_3)/n_B$ as a parameter. If the isospin
composition is conserved in addition to the baryon density $n_B$,
then $w$ is also a constant.

\begin{figure}[t]
    \centering
    \includegraphics[height=0.4\textheight] {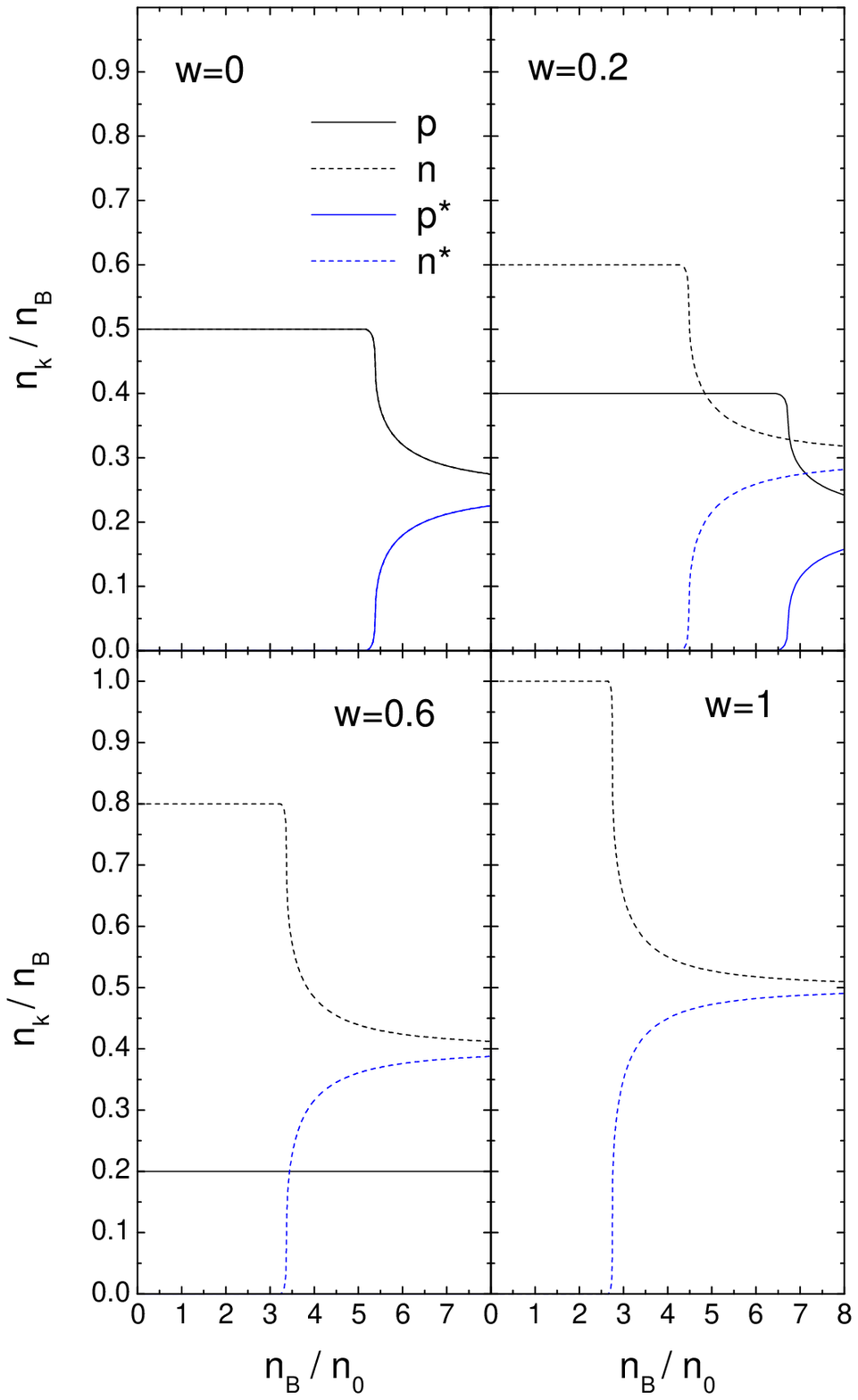}
    \caption{The population of different baryons in dense matter in terms
    of the total baryonic density for
   zero temperature and different isospin fractions,.}
    \label{IsoPop}%
\end{figure}

As in the last section, it is found that low density matter is
composed only by protons and neutrons. But as the density is
increased, their parity partners become active as it is shown in
Fig. \ref{IsoPop}. This figure corresponds to zero temperature and
includes several isospin fractions. For isospin symmetric matter
($\text{w}=0$) protons and neutrons are indistinguishable, as well
as $p^*$ and $n^*$ are. By increasing the asymmetry, a splitting
of the partners population is found. In addition, the onset of the
$n^*$ is shifted downward, while that of $p^*$ is raised to higher
densities. Thus, for $\text{w}=0.6$ the $p^*$ does not appear
within the range analyzed in this figure and, consequently, its
partner $p$ shows a constant trend. For pure neutron matter
($\text{w}=1$), the $n$ and $n^*$ have a specular behavior and
asymptotically tend to an equipartition. This is a manifestation
of the chiral recovery, since as the masses of the partners tends
to the common value $m_0$, they become statistically
indistinguishable.\\
\begin{figure}[h]
    \centering
    \includegraphics[height=0.36\textheight] {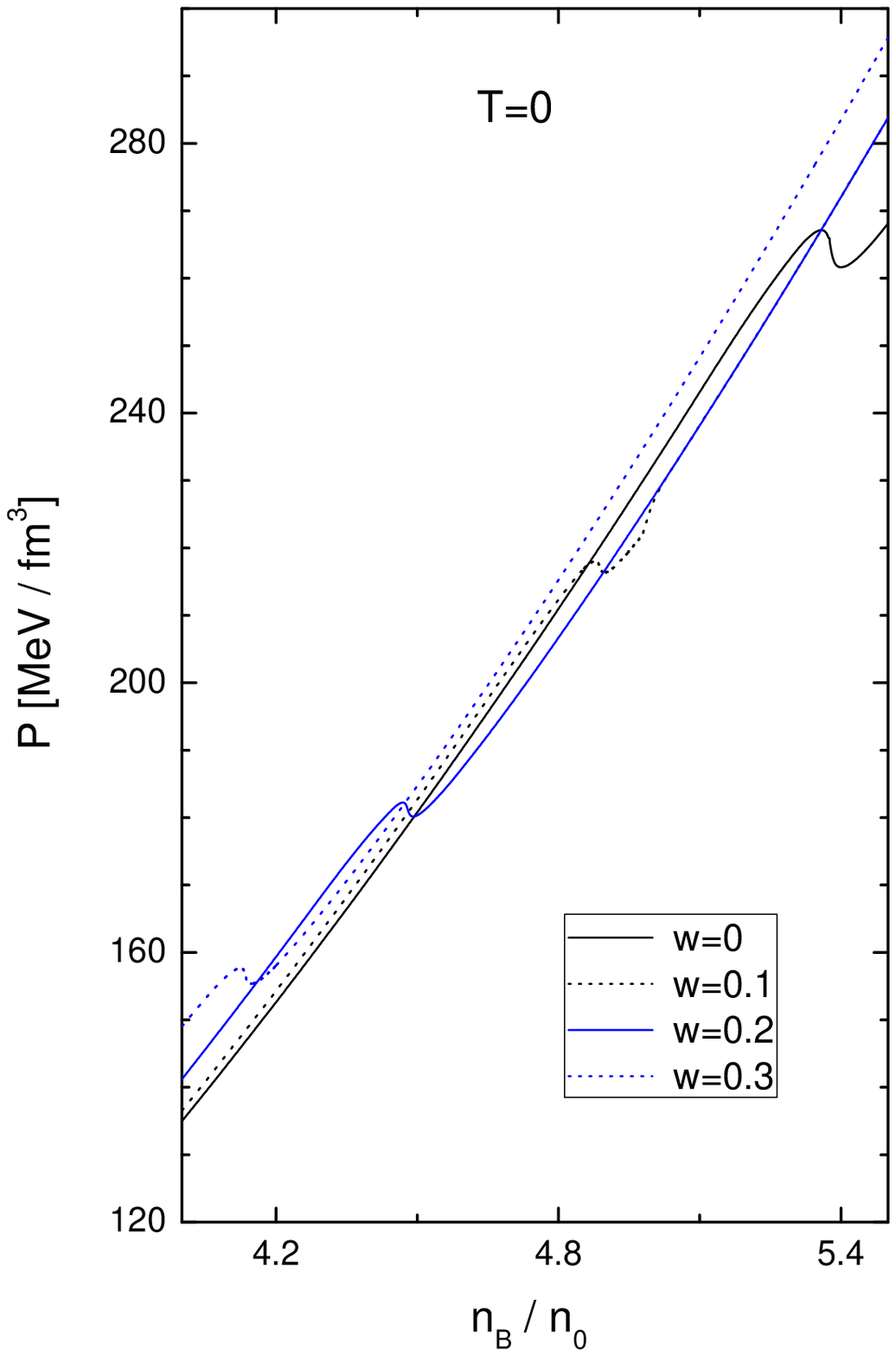}
    \caption{The pressure in terms of a limited range of densities
    and several isospin asymmetries at zero temperature.}
    \label{AsymPress}%
\end{figure}

The emergence of the dual partners leaves its imprints in the
equation of state, as shown in Fig. \ref{AsymPress}. For each of
the isospin compositions considered here, the pressure is
monotonically increasing until the presence of the $n^*$ is strong
enough to cause a first order phase transition. At this point the
pressure shows a typical van der Waals behavior concentrated in a
narrow range of density, with its unstable and metastable
sections. It can be appreciated that a small change in the
asymmetry drastically changes the location of the instability, for
this reason only a limited range for $\text{w}$ is shown in this
figure. The location of the critical density progressively
decreases with $\text{w}$, from $n_B/n_0 \simeq 5.3$ for symmetric
matter it falls to $n_B/n_0 \simeq 2.7$ for pure neutron matter.
\begin{figure}[b]
    \centering
    \includegraphics[height=0.4\textheight] {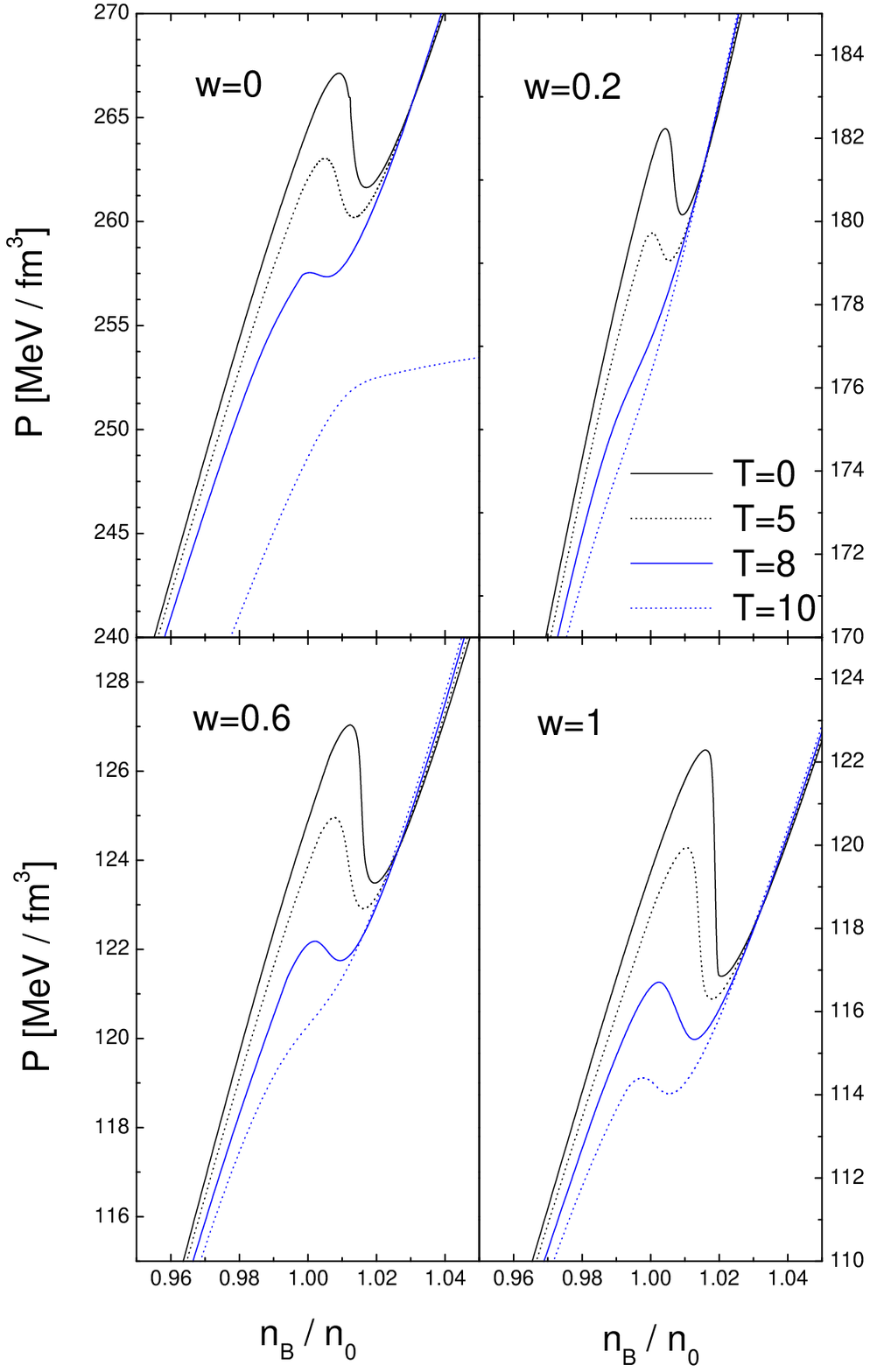}
    \caption{The pressure in terms of the density, limited to the
    range of the instability. Several isospin asymmetries and temperatures
    are considered.}
    \label{AsymPTw}%
\end{figure}

To take account of the thermal effects on the pressure, the
instability domain is shown for several asymmetries and
temperatures in Fig. \ref{AsymPTw}.
For a given isospin fraction the pressure curves become flatter as
the temperature increases, until a maximum value $T_m$ where they
acquire a monotonously increasing trend. This figure shows that
$T_m$ has a dependence on $\text{w}$, for instance for
$\text{w}=0.2$ is $ T_m \simeq 5$ MeV, then it increases for
$\text{w}=0$ and $0.6$ within the range $8 < T_m  < 10$, and
finally it exceeds $T=10$ MeV for $\text{w}=1$.
\begin{figure}[h]
    \centering
    \includegraphics[height=0.36\textheight] {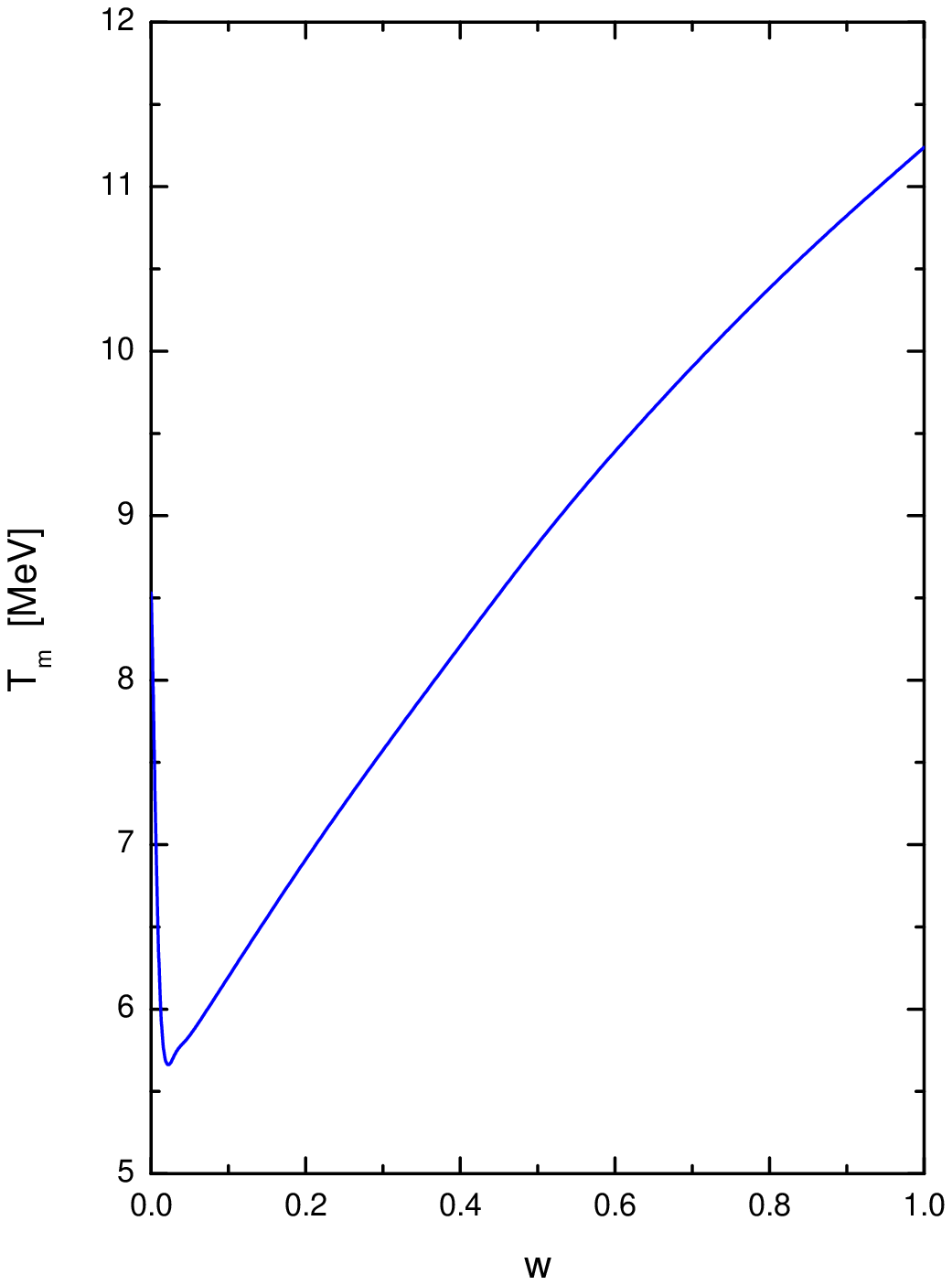}
    \caption{The maximum temperature for the first order chiral transition in terms of the
    isospin asymmetry.}
    \label{Tmax}%
\end{figure}
Two conclusions can be drawn from this figure, the critical
temperature for this first order phase transition should take
place for the maximum asymmetry, furthermore the temperature $T_m$
takes on a minimum value for nonzero but low asymmetries. The
specific details are shown in Fig. \ref{Tmax}, where $T_m$ is
shown in terms of $\text{w}$. A sudden drop of almost $30 \%$ is
found for matter with a slight neutron excess $\text{w}=0.02$ as
compared with the perfectly isospin symmetric case. Up from this
point, this characteristic temperature grows uniformly and reaches
its maximum value $T_c=11.2$ MeV for pure neutron matter.\\
The comparison with published results using the DCPM is adequate
at this point, although the criterium for the selection of
parameters is not strictly coincident. For instance the maximum
temperature for the first order chiral phase transition at
$\text{w}=0$ is near $T_m \sim 10$ MeV in \cite{Motohiro}, and
$T_m=7$ MeV in \cite{Koch}. The critical density under the same
conditions, varies within $5.9 < n_B/n_0 < 7.9$ when the parameter
$m_0$ ranges from $m_0=600$ MeV to $m_0=700$ MeV. In the present
work a slightly lower value $n_B/n_0=5.3$ is obtained for the set
B, which has a parameter $m_0$ within the mentioned interval, see
Table \ref{Table1}. A very different result  $n_B/n_0=7.8$ is
obtained in \cite{Koch}, notwithstanding it is consistent with the
use of a relatively high $m_0=750$ MeV.

\begin{figure}[h]
    \centering
    \includegraphics[height=0.4\textheight] {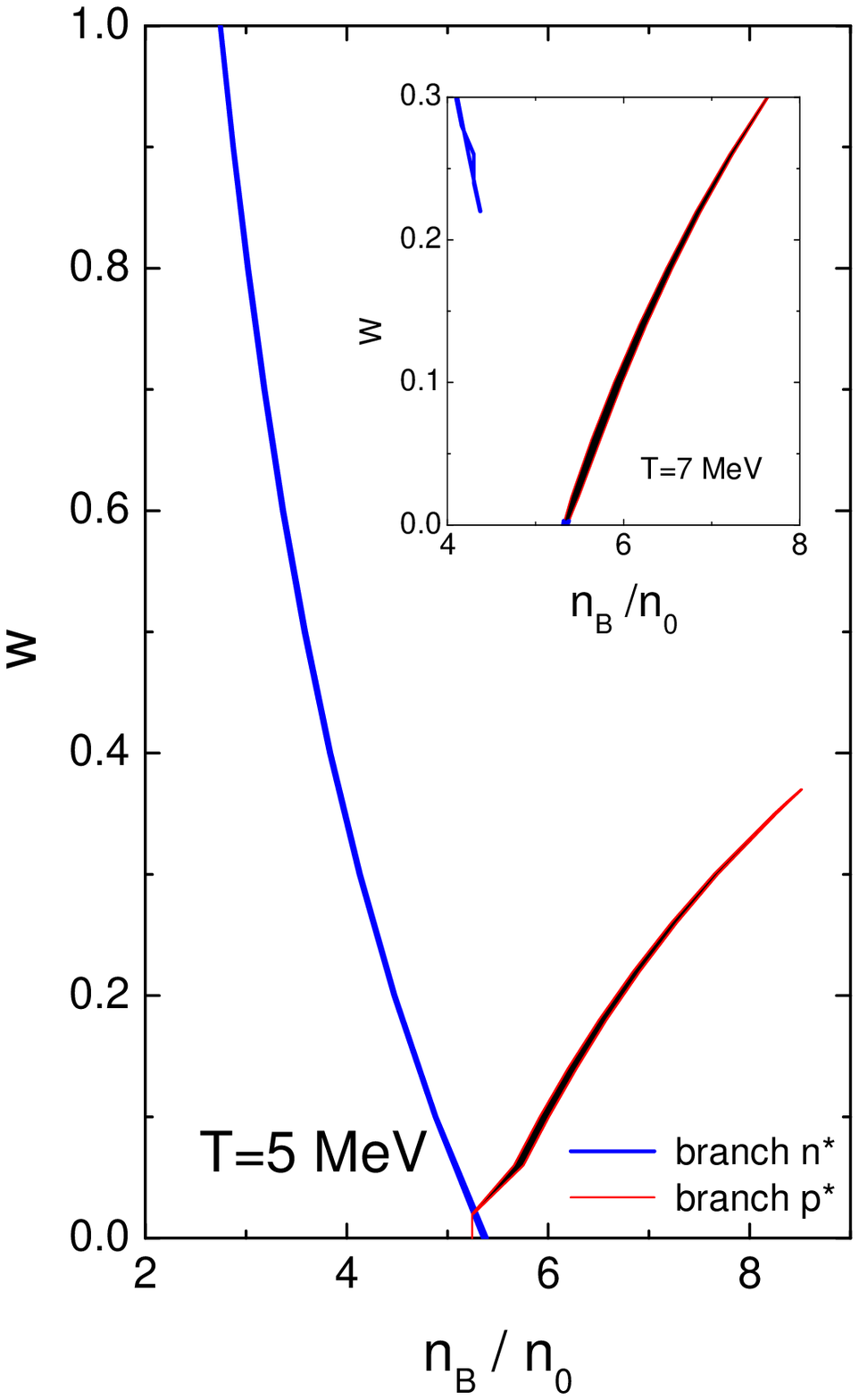}
    \caption{The spinodal surface in the $n_B - \text{w}$ plane for two characteristic temperatures.
    In order to show its full extension, a scale is used for which the surface seems like a thin
    curve.}
    \label{Unstable}%
\end{figure}

\begin{figure}[h]
    \centering
    \includegraphics[height=0.4\textheight] {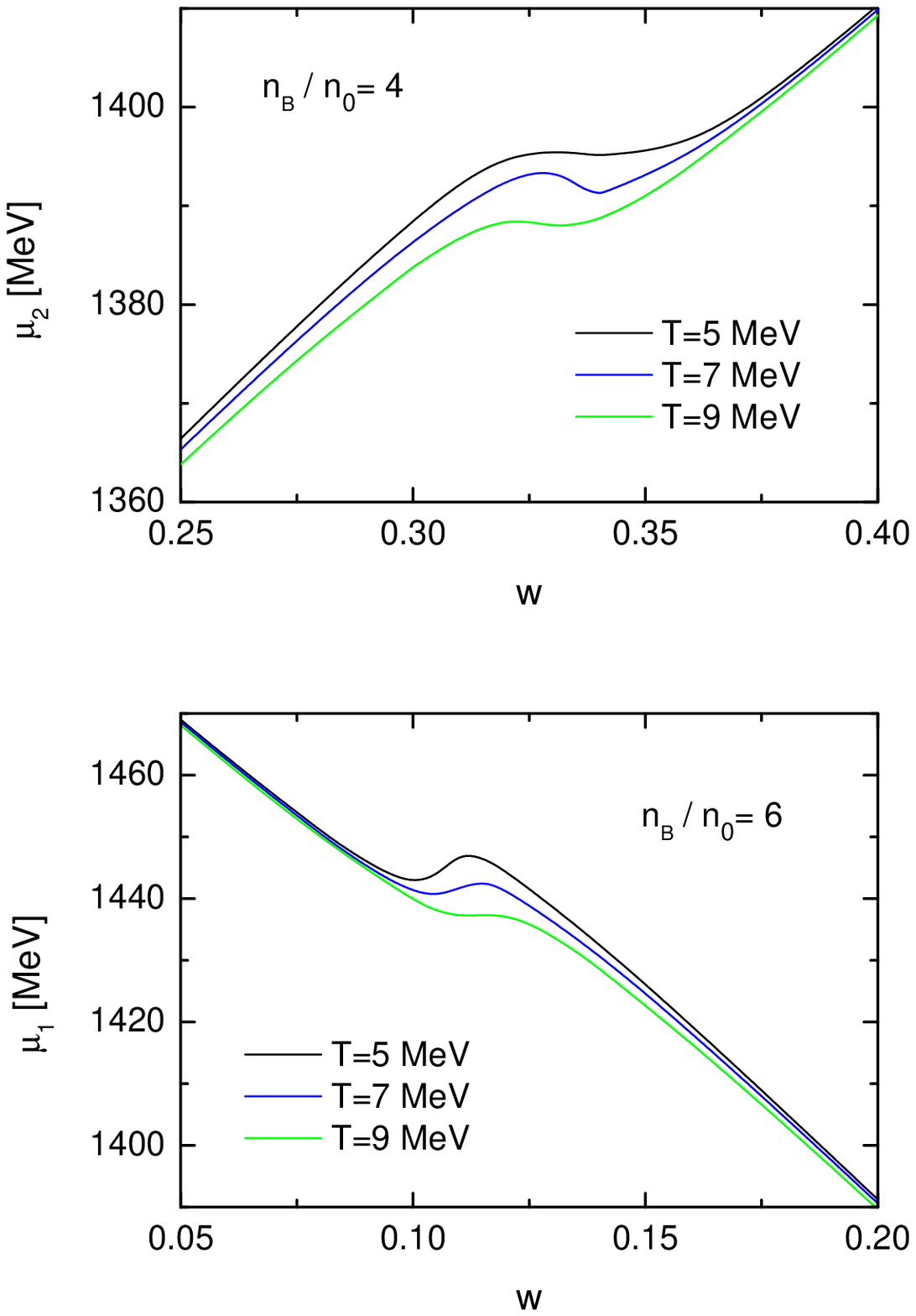}
    \caption{The chemical potential of the neutron (proton) in the upper (lower)
    panel in terms of the isospin asymmetry for two representative
     densities and several temperatures. The non-monotonous trend is a signal of
     the violation of the chemical equilibrium.}
    \label{CECond}%
\end{figure}

The phase transition is the consequence of the violation of the
thermodynamical equilibrium criteria, which can be stated in terms
of the change of convexity of the free energy ${\cal
F}(T,n_B,n_I)$. The conditions
\begin{equation} \frac{\partial^2
{\cal F}}{\partial n_B^2}>0,\;\; \frac{\partial^2 {\cal
F}}{\partial n_B^2} \,\frac{\partial^2 {\cal F}}{\partial n_I^2} -
\left(\frac{\partial^2 {\cal F}}{\partial n_B \partial
n_I}\right)^2 >0 \label{ThermodInst} \end{equation}
guarantees the stability of a bi-component system
\cite{Muller,BaoLi}.  They are usually known as the mechanical and
chemical stability conditions respectively. In Fig. \ref{Unstable}
these relations have been used to map the instability region,
which is bounded by the spinodal curve.   In the main panel the
case $T=5$ MeV is representative for temperatures below $T\simeq
6.5$ MeV. It is evident that unstable matter can be found for the
whole range $\text{w}>0$, and temperatures below $T_c$.
Furthermore, the phase transition is sharply focused in the phase
space as implied by the fact that the spinodal surface seems like
a thin curve for the scale used. There are two separate branches,
the low density one spreads over the full range $0 < \text{w} <1$,
and corresponds to the violation of both mechanical and chemical
equilibrium. It is associated with the emergence of the neutral
partner $n^*$, since the spinodal coincides with the threshold
where this particle reaches a concentration of about $5 \%$. The
charged partner $p^*$, instead, has a negligible presence for $0<
w < 0.3$ or it is missing at all for $\text{w} > 0.3$. In
contrast, the high density branch only breaks the chemical
equilibrium but fulfills the mechanical condition. It marks the
point where the particle $p^*$ becomes thermodynamical relevant.
It extends only up to $w=0.3$ because at this temperature the
incoming of $p^*$ is disfavored for greater asymmetries.\\
A slight variation of this situation is found as the temperature
is raised, as shown in the inserted panel in the same figure. In
this case the low density branch splits into separate parts, one
of them exists for $\text{w} > 0.22$ because $T_m < 7$ MeV for
$\text{w} < 0.22$ (see Fig. \ref{Tmax}). The other part is
restricted to almost symmetric matter $\text{w} <0.003$ and in
this figure it could be mistaken as the starting point of the high
density branch. The gap
in $\text{w}$ separating both parts increases with the temperature.\\
The existence of two disjoint branches of unstable matter,
spreading over very different domains of the phase space, remarks
that the information provided by the equation of state of Fig.
\ref{AsymPress} is insufficient to sketch a panoramic view in the
present model. In order to improve this issue, the behavior of the
chemical potentials in terms of the isospin asymmetry is shown in
Fig. \ref{CECond}. For this purpose two representative baryon
densities are chosen, in the upper panel the case $n_B/n_0=4$ is
affected by the low density branch, which is characterized by both
mechanical instability and a non-monotonous increase of $\mu_2$
with the asymmetry. On the lower panel, instead, the feature of
the high density branch is exhibited for $n_B/n_0=6$. In this case
the loss of thermodynamical equilibrium manifests in the irregular
decrease of $\mu_1$ with $\text{w}$, since the onset of $p^*$
leads this instability as mentioned earlier.

A complete study of asymmetric matter within the CDPM was given in
\cite{EserBlaiz2}, using an approach similar to the present one.
The main differences are the use of the $a_0(980)$ meson and the
lack of  the Dirac sea effects in the present work. The
regularization at zero temperature made in \cite{EserBlaiz2}
comprises the tree level divergence of the grand potential in Eq.
\ref{GrandP}, and also the vertices $C_0$ and $C_1+C_2$, which are
beyond that order. The comparison of the spinodal surface in the
w-$n_B$ plane shows that the effect of the scalar isovector meson
is really significant. Furthermore, the critical temperature in
\cite{EserBlaiz2} is $T_c=8.5$ MeV and it takes place for
symmetric matter. In contrast, a grater value $T_c=11.2$ is
obtained here, and fundamentally it occurs for pure neutron
matter.

Isospin symmetric matter  as well as pure neutron matter play a
special role, since in both instances there is only one chemical
potential for all the particles. In the first case both branches
of instability coincide, and in the latter case the proton and its
chiral partner are excluded by definition, eliminating the high
density branch. However, for both of them the initial and final
points of the phase transition preserve the isospin asymmetry. It
is interesting to study the energy balance under such conditions.
It is well known that a first order phase transition gives rise to
a latent heat, which for an isothermal process at constant
chemical potential can be evaluated as
\[ \ell =T\left(\;\;\left[\frac{{\cal S}}{n_B}\right]_b- \left[\frac{{\cal S}}{n_B}\right]_a\right), \]
the subindexes represent the extreme states of the transition, and
by convention $a$ stands for the denser one. The results displayed
in Fig.  show the typical bell-shaped curves, with a scale of the
order of keV, several orders of magnitude lesser than the results
commonly obtained for the liquid-gas phase transition. This is in
accordance with the fact that the spinodal has a considerably
reduced extension in the phase space.

\section{SUMMARY AND CONCLUSIONS}\label{Colussions}

In this work the possible manifestations of the chiral symmetry in
dense nuclear matter have been studied, using a model of hadronic
fields which realise the chiral symmetry under the mirror
assignment \cite{Detar,Jido}. For this purpose the CDPM introduces
the parity partners $n,\, n^*$ and $p,\, p^*$. The parameters of
the model have been been determined in the MFA, imposing
conditions over a wide range of matter densities. The model
adjusts the masses in vacuum of the nucleons and their parity
partners as well as those of the pions and the low lying scalar
mesons $\sigma$ and $a_0(980)$. It also describes the key features
of nuclear matter at the saturation density and zero temperature.
The highest densities have been included through the requirement
of predicting a mass-radius relation for neutron stars that is
compatible
with the recent analysis \cite{Miller,Luo2024,Vincig}.\\
The focus has been posed on isospin asymmetric matter at finite
termperature, therefore the model includes the interaction with
the vector isovector $\rho$ and scalar isovector $\zeta$ mesons,
the latter identified as the $a_0(980)$. Initially, a coupling
between vector mesons was proposed (Eq. \ref{LagVec}), including a
$\omega-\rho$ mixing vertex. However, it has been found that the
reproduction of the phenomenological
constraints does not favor its inclusion.\\
The emergence of the parity partners to the Fermi sea produces
thermodynamical instabilities which lead to first order phase
transition. These instabilities are described by two separate
branches, the low density one has the strongest effects and
manifests by a non-monotonous behavior of the equation of state.
It has a critical temperature $T_c=11.2$ MeV. The other branch is
generated by diffusive instabilities associated with the emergence
of $p^*$, and takes place for higher densities. The spinodal
extends over the full range of isospin asymmetries $0\leq \text{w}
\leq 1$, although it is extremely concentrated. Instabilities of
the same characteristics have been found even for neutron star
matter. The coupling to the $\zeta$ meson  is the main cause of
such drastic behavior, since its coupling to the nucleons
distinguish the isospin structure. Therefore a splitting of the
effective nucleon masses occurs, giving a different dynamical
response for each member of the isospin duplet. This effect
reinforces the difference in the energy spectrum due to the
coupling to the vector isovector meson. However, the influence of
the $\rho$ meson must be of lesser magnitude since this phenomenon
has not been reported in previous investigations which only
include the vector meson in their conceptual framework to describe
isospin imbalanced dense matter. Thus, it must be concluded that a
complete treatment of the chiral transition under such conditions
should not do without the coupling of the
nucleons to the scalar isovector meson. \\
The specific latent heat, evaluated for symmetric nuclear matter
and pure neutron matter, shows that the energy released by a
particle participating in the chiral phase transition is of the
order of the keV. This result is significantly lower than other
characteristic transitions, as for instance the liquid-gas
conversion at low density.

There are interesting issues, not considered in this work, which
will be object of further investigation such as the effect of
vacuum contributions or the possible consequences of the chiral
phase transition on the dynamics of neutron stars.

\section*{Acknowledgements}
This work has been partially supported by CONICET, and by
Universidad Nacional de La Plata, Argentina.

\bibliography{Paper}

\end{document}